\documentclass[12pt]{article}
\usepackage[left= 0.8in, textwidth=6.9in, textheight=9.0in, top=1.0in]{geometry}
\usepackage{amssymb,amsmath}
\usepackage{graphicx}

\newcommand{\MT}{\left[ \begin{array}{rrrrrrrrrrrrrrrrrrrr}}
\newcommand{\EM}{\end{array}\right]}
\newcommand{\EQ}{\begin{equation}\begin{array}{lllllllllll}}
\newcommand{\EE}{\end{array}\end{equation}}
\newcommand{\Real}{\mathbb R}
\newcommand\norm[1]{\left\lVert#1\right\rVert}
\newcommand\abs[1]{\left|#1\right|}

\def\bff{{\bf f}}

\def\bfh{{\bf h}}
\def\bfx{{\bf x}}
\def\bfy{{\bf y}}
\def\bfe{{\bf e}}
\def\bfu{{\bf u}}
\def\bfz{{\bf z}}
\def\bfW{{\bf W}}
\def\bfb{{\bf b}}

\def\ds{\displaystyle}
\def\Fr{\ds \frac}

\title{\LARGE \bf
The Observability in Unobservable Systems
}
\date{}

\author{Wei Kang$^{1}$, Liang Xu$^{2}$ and Hong Zhou$^{3}$% <-this % stops a space
\thanks{$^{1}$Wei Kang, Department of Applied Mathematics, Naval Postgraduate School, Monterey, California \& University of California, Santa Cruz, California, USA
        {\tt\small wkang@nps.edu}}%
\thanks{$^{2}$ Liang Xu, Marine Meteorology Division, Naval Research Laboratory, Monterey, California, USA {\tt\small liang.xu@nrlmry.navy.mil}}%
\thanks{$^{3}$ Hong Zhou, Department of Applied Mathematics, Naval Postgraduate School , Monterey, California, USA {\tt\small hzhou@nps.edu}}%
}

\begin{document}

\maketitle
%\thispagestyle{empty}
%\pagestyle{empty}

%%%%%%%%%%%%%%%%%%%%%%%%%%%%%%%%%%%%%%%%%%%%%%%%%%%%%%%%%%%%%%%%%%%%%%%%%%%%%%%%
\begin{abstract}

In this paper, we introduce the concept of observability of targeted state variables for systems that may not be fully observable. For their estimation, we introduce and exemplify a deep filter, which is a neural network specifically designed for the estimation of targeted state variables without computing the trajectory of the entire system. The observability definition is quantitative rather than a yes or no answer so that one can compare the level of observability between different sensor locations. 

\end{abstract}

%%%%%%%%%%%%%%%%%%%%%%%%%%%%%%%%%%%%%%%%%%%%%%%%%%%%%%%%%%%%%%%%%%%%%%%%%%%%%%%%
\section{INTRODUCTION}

In this paper, we study the observability and estimation methods of targeted state variables for systems that are not necessarily observable. In control theory, the observability is a widely used concept that defines the feasibility of estimating the state, or initial state, of dynamical systems based on their outputs. In this paper, we address the problem of partial observability. If a system is not observable, is it possible to take the advantage of its output to estimate a part of the state variables? Even if a system is observable, is that possible to estimate only the targeted state variables without using a full scale filter that estimates all variables in the state space? The answer to these questions is important for various reasons. The ever increasing complexity and dimension of dynamical systems in science and engineering result in many problems for which the system is either unobservable or computationally too expensive to provide a real-time estimation of the entire system states. Some examples of such systems include numerical weather prediction and large swarms of unmanned vehicles. In \cite{otsuka} and \cite{shi}, for instance, deep learning is applied to nowcasting, i.e., predicting the future rainfall intensity in a local region over a relatively short period of time without running a full-scale data assimilation system. In \cite{gong}, it is proved that partially estimating some parameters in swarms of unmanned vehicles is possible when the overall system is unobservable. 

There is a huge literature on the theory of nonlinear estimation and filtering. It is worth to note that deep learning as a tool of nonlinear filtering, namely deep filter, is analyzed in \cite{zhang}. We adopt a similar approach in Section \ref{sec_deepfilter}. Different from the problem addressed in \cite{zhang} where the system is observable, we use deep filters to approximate a targeted individual variable in a system that may not be fully observable. Measuring observability quantitatively based on the observabililty Gramian is essential to the type of problems studied in this paper. For related literature, the interested readers are referred to \cite{gong,kailath,kangxu, kangxu1,kangxu2,kangxu3,krener,powel,qi,qi1} and references therein. 

In this paper, we introduce the concept of observability of targeted state variables. For their estimation, we introduce and exemplify a deep filter, which is a neural network specifically designed for the estimation of targeted state variables without computing the trajectory of the entire system.

\section{Observability}
Consider a discrete-time system
\EQ
\label{eq_sys}
\bfx(k+1)=\bff(\bfx (k)),\\
\bfy (k)=\bfh(\bfx (k)),
\EE
where $k\geq 0$ is an integer and 
\EQ
\bfx=\MT x_{1}&x_{2}&\cdots&x_{n}\EM^\intercal\in\Real^n\\
\bfy=\MT y_{1}&y_{2}&\cdots&y_{m}\EM^\intercal\in\Real^m
\EE 
are the state and output variables, respectively. Given $\bfy(0),\bfy(1),\cdots,\bfy(K)$, where $K\geq 1$ is an integer, we define the observability of an individual state. Without loss of generality, let us consider the observability of $x_1(K)$. It is worth to note that we measure the observability of the final state at $k=K$. However, similar ideas in this section can be applied to the observability of the initial state $x_1 (0)$. In general, the quantitative measure of observability has different value for the initial and final states. Following the general definition of observability introduced in \cite{kangxu}, we define the observability of  $x_1(K)$. Consider a trajectory, $\{ \bfx (k); 0\leq k\leq K\}$, and the following problem of constrained maximization,
\EQ
\label{def1}
\rho^2=\ds\max_{\hat\bfx(0),\cdots,\hat\bfx(K)} \{ (\hat x_1(K)-x_1(K))^2\} \\
\hspace{-0.2in} \mbox{subject to}\\
\ds\sum_{j=0}^K\norm{\hat\bfy(k)-\bfy(k)}_2^2 \leq \epsilon^2\\
\hat \bfx(k+1)=\bff(\hat \bfx(k)),\;\;\;\;\;\;\; k=0,1,\cdots,K-1\\
\hat \bfy(k)=\bfh(\hat \bfx(k)), \,\;\;\;\;\;\;\;\;\;\;\;\; k=0,1,\cdots,K
\EE
where $\epsilon>0$ is a constant representing the upper bound of output variations and  $0\leq\rho\leq \infty$ is called the ambiguity in the estimation of $x_1(K)$. The ratio $\rho/\epsilon$ is called the {\it unobservability index}, a term adopted from \cite{krener}. If the output variable represents the sensor information, then it has measurement error. The true value of $\bfy(k)$ is unknown. What we know is that the truth should be in a neighborhood of the measured data. Therefore, we allow in (\ref{def1}) the output to take any value in an $\epsilon$-neighborhood. The trajectory $\{ \hat \bfx(k); 0\leq k\leq K\}$ in (\ref{def1}) that has the largest variation of $\abs{\bar x_1(K)-x_1(K)}$ is considered the worst possible estimation of $x_1(K)$ whose output is within an $\epsilon$-neighborhood of the measured $\bfy$. To summarize, if the value of the unobservability index, $\rho/\epsilon$, is small, then $x_1(K)$ is observable. The range of the unobservability index value for observable systems depends on the scale of the uncertainties in $\bfy$.  In this paper, we use this quantitative measure of observability to find individual variables that can be estimated accurately and to compare the observability between different sensor locations. 

Consider a linear system,
\EQ
\bfx(k+1)=A\bfx(k);\\
\bfy(k)=H\bfx(k)
\EE
where $A\in\Real^{n\times n}$ and $H\in \Real^{m\times n}$. Then (\ref{def1}) is equivalent to
\EQ
\label{def2}
\rho^2=\ds\max_{\varDelta \bfx_0} \varDelta \bfx_0^\intercal F^\intercal F\varDelta \bfx_0,\\
\hspace{-0.2in} \mbox{subject to}\\
\varDelta \bfx_0^\intercal G \varDelta \bfx_0=\epsilon^2,
\EE
where $\varDelta \bfx_0$ represents $\hat \bfx_0-\bfx_0$ and
\EQ
\label{def2b}
G=\ds\sum_{k=0}^K (A^\intercal)^kH^\intercal HA^k,\\
F=\MT 1 &0\cdots 0\EM A^K.
\EE
The computation of the unobservability index boils down to numerically solving the constrained quadratic maximization (\ref{def2})-(\ref{def2b}). The matrix $G$ is, in fact, the observability Gramian, a symmetric matrix that is used to measure observability in control theory \cite{kailath,kangxu}. For problems that have high dimensions, numerically solving (\ref{def2}) is not straightforward if the overall system is weakly observable or unobservable because, in this case, the condition number of $G$ can be very large. Techniques of computational linear algebra have to be applied when solving (\ref{def2}). Due to length limitations, these techniques will be addressed in another paper. 

For a nonlinear system, the computation is more difficult because (\ref{def1}) is a nonconvex problem, which is not easy to solve. On the other hand, we can approximate the observability by solving (\ref{def2}) based on the linearization of the nonlinear system along nominal trajectories. Specifically, $G$ and $F$ can be computed empirically using the central difference method. Let $\delta>0$ be a small number. Around a trajectory $\{ \bfx(k); 0\leq k\leq K\}$, we compute $2n$ trajectories 
\EQ
\label{eq_empirical1}
\bfx^{\pm i}(0)=\bfx(0)\pm \delta \bfe_i, \\
\bfx^{\pm i}(k+1)=\bff(\bfx^{\pm i}(k)), & k=0,1,\cdots,K-1\\
\bfy^{\pm i}(k)=\bfh(\bfx^{\pm i}(k)), & k=0,1,\cdots, K
\EE
for $i=1,2,\cdots,n$, where $\bfe_i$ is the $i$th unit vector in $\Real^n$.  Define 
\EQ
\label{eq_empirical2}
\varDelta^i\bfy(k)=( \bfy^{+i}(k)-\bfy^{-i}(k))/(2\delta), \\
\varDelta\bfy(k)=\MT \varDelta^1\bfy(k)&\varDelta^2\bfy(k)&\cdots&\varDelta^n\bfy(k)\EM, \\
\varDelta^{ i}x_1(K)=(x^{+i}_1(K)-x^{-i}_1(K))/(2\delta).\\
\EE
Then, $G$ and $F$ are approximated by
\EQ
\label{eq_empirical3}
G=\ds\sum_{k=0}^K \varDelta\bfy(k)^\intercal \varDelta\bfy(k),\\
F=\MT \varDelta^{1}x_1(K) & \varDelta^{2}x_1(K)&\cdots&\varDelta^{n}x_1(K)\EM
\EE
In the following, the value of $\rho/\epsilon$ is approximated by solving (\ref{def2}) using the empirical approximation of $G$ and $F$ in (\ref{eq_empirical3}). This approach avoids the requirement of solving the nonconvex optimization problem (\ref{def1}). 

\section{Estimating state variables of unobservable systems}
\label{sec_UKF}
\begin{figure}[!ht]
\centering
\includegraphics[width = 3.0in]{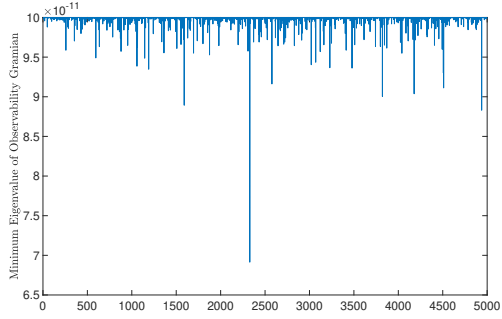}
\caption{ The minimum eigenvalue of the observability Gramian at $5,000$ sample points.}
\label{fig_eign_G}
\end{figure}

In this section, we use an example to illustrate the idea of estimating a targeted state variable when the overall system is unobservable. Consider Burgers' equation
\EQ
\label{burgers}
\ds\frac{\partial U(x,t)}{\partial t} +U(x,t)\ds\frac{\partial U(x,t)}{\partial x} =\kappa \ds\frac{\partial^2 U(x,t)}{\partial x^2}, \\
U(x,0)=U_0(x),\\
U(0,t)=0, \;U(L,t)=0.
\EE
where  $(x,t)\in [0, L]\times [0, T]$. The solution is approximated by solving a finite dimensional  discretized system. The discretization is based on central difference in space and $4$th order Runge-Kutta in time (see, for instance, \cite{kangxu2}). Let $N_t>0$ and $N_x>0$ be integers. The discretized trajectory is represented as follows,

\EQ
\label{eq_discretate_traj}
\bfu(k)=\MT u_1(k),u_2(k),\cdots, u_{N_x-1}(k)\EM^\intercal,
\EE
for $k=0,1,\cdots, N_t$. In (\ref{eq_discretate_traj}), $u_i(k)$ represents $U(x,t)$ evaluated at the grid point $(x_i,t_k)$, where
\EQ
x_i=i\varDelta x, & i=1,2,\cdots, N_x-1\\
t_k=k\varDelta t, & k=0,1,\cdots, N_t,\\
\varDelta x=L/N_x,\\ 
\varDelta t=T/N_t.
\EE
Note that the boundaries $u_0(k)=0$ and $u_{N_x}(k)=0$ are known. In this example, we set $T=5$, $L=2\pi$, $\kappa=0.14$, $N_x=50$ and $N_t=100$. For the output, we assume that sensors are located at $x_{20}$, $x_{21}$, $x_{29}$ and $x_{30}$, i.e., 
\EQ
\label{eq_burger_output}
\bfy(k)=\MT u_{20}(k)& u_{21}(k) & u_{29}(k)& u_{30}(k)\EM^\intercal
\EE
For nonlinear systems, the observability depends on the location of the trajectory. For analysis, we take a data-driven approach. We randomly select $5,000$ initial states around $U(x,0)=0$. Their empirical Gramians are computed using (\ref{eq_empirical1})-(\ref{eq_empirical3}) for $K+1=10$, i.e. measurements of $10$ time steps of $\bfy$ are used in the state estimation. The smallest eigenvalue of $G$ at the samples is shown in Figure \ref{fig_eign_G}. The eigenvalues are small, all around $10^{-10}$. This implies that the observability of the system is extremely weak, or the system is practically unobservable. 

\begin{figure}[!ht]
\centering
\includegraphics[width = 3.0in]{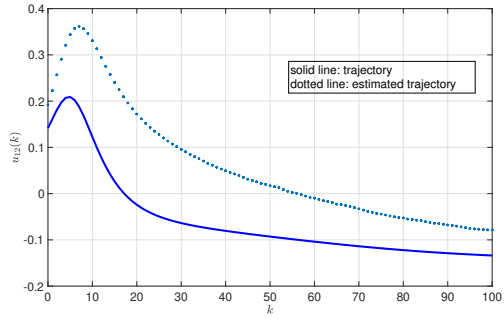}
\caption{ The true value of $u_{12}(k)$ and its UKF estimation. The state $u_{12}(K)$ is practically unobservable because $\rho/\epsilon = 1.1884\times 10^{4}$.}
\label{fig_err_u12}
\end{figure}

Unscented Kalman filter (UKF) \cite{julier} is applied to the system. Because the observability is extremely weak, we do not expect an accurate estimation of all state variables. In the simulation, we apply  i.i.d. random noise  to $\bfy$. The standard deviation is $\sigma = 0.028$. Due to the extremely weak observability, the filter cannot correct the initial estimation error. One example  is shown in Figure \ref{fig_err_u12}. The initial error in the estimation of $u_{12}$ cannot be corrected efficiently. 

\begin{figure}[!ht]
\centering
\includegraphics[width = 3.0in]{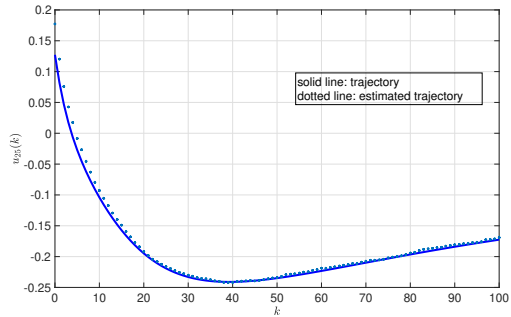}
\caption{ The true value of $u_{25}(k)$ in a trajectory and its UKF estimation.}
\label{fig_err_u25a}
\end{figure}

\begin{figure}[!ht]
\centering
\includegraphics[width = 3.0in]{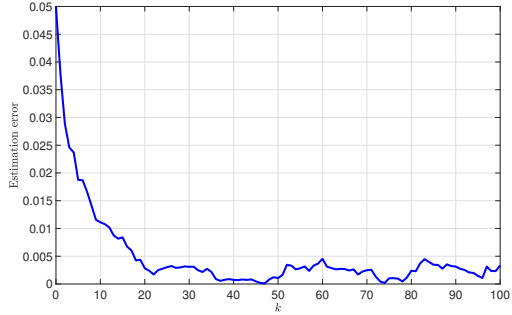}
\caption{ The error of UKF estimation $\abs{\hat u_{25}(k)-u_{25}(k)}$.}
\label{fig_err_u25b}
\end{figure}

Although the overall system is unobservable, some individual state variables can still be reasonably observable under the same output function (\ref{eq_burger_output}). For example, by solving (\ref{def2}) using the empirical Gramian (\ref{eq_empirical3}), the unobservability index of $u_{25}(K)$ is $\rho/\epsilon = 4.3702$ if $K+1=10$ and $\rho/\epsilon = 0.7940$ if $K+1=20$. This is in sharp contrast to the unobservable state $u_{12}(K)$ shown in Figure \ref{fig_err_u12}, for which $\rho/\epsilon = 1.1884\times 10^{4}$. What does this mean? Assume that the sensor error of each measurement is $0.028$. The total $l_2$ norm of sensor error for $4\times 10$ measurements ($K+1=10$) is about $\norm{e}_2=0.1771$ and $\norm{e}_2=0.2504$ for $4\times 20$ measurements ($K+1=20$). Multiplying $\norm{e}_2$ by the value of $\rho/\epsilon$, it implies that the worst error in the estimation of $u_{25}(K)$ is about $0.7740$ when $K+1=10$ and $0.1988$ when $K+1=20$. They may not seem to be very small. But we would like to emphasize that they are the worst possible error. An estimator, such as UKF, that optimizes the estimation avoids the worst scenario. In fact, the UKF estimation of $u_{25}(k)$ has a much smaller error than the worst case. Shown in Figures \ref{fig_err_u25a}-\ref{fig_err_u25b} , the UKF estimation of $u_{25}(k)$ converges to the true value quickly; the initial error is reduced by more than 90\% after $k=20$ steps. 

\section{Deep filter}
\label{sec_deepfilter}
A UKF filter estimates all state variables even if the goal is to find an estimation of a single targeted variable. This is inefficient. In fact, for some high dimensional systems, a real-time estimation of the entire system is simply impossible due to the high computational load. In this section, we introduce a deep filter. The method is based on deep learning. An advantage of a deep filter is that the computation is focused on the targeted state variable to be estimated, without the requirement of running a full scale filter that estimates all state variables simultaneously. In \cite{zhang}, deep neural networks are applied to approximate state variables based on the output. In this section, we apply a similar idea. But different from \cite{zhang} where the system is observable, we use deep neural networks to approximate an individual variable in a system that may not be fully observable. 

\begin{figure}[!ht]
\centering
\includegraphics[width = 2.5in]{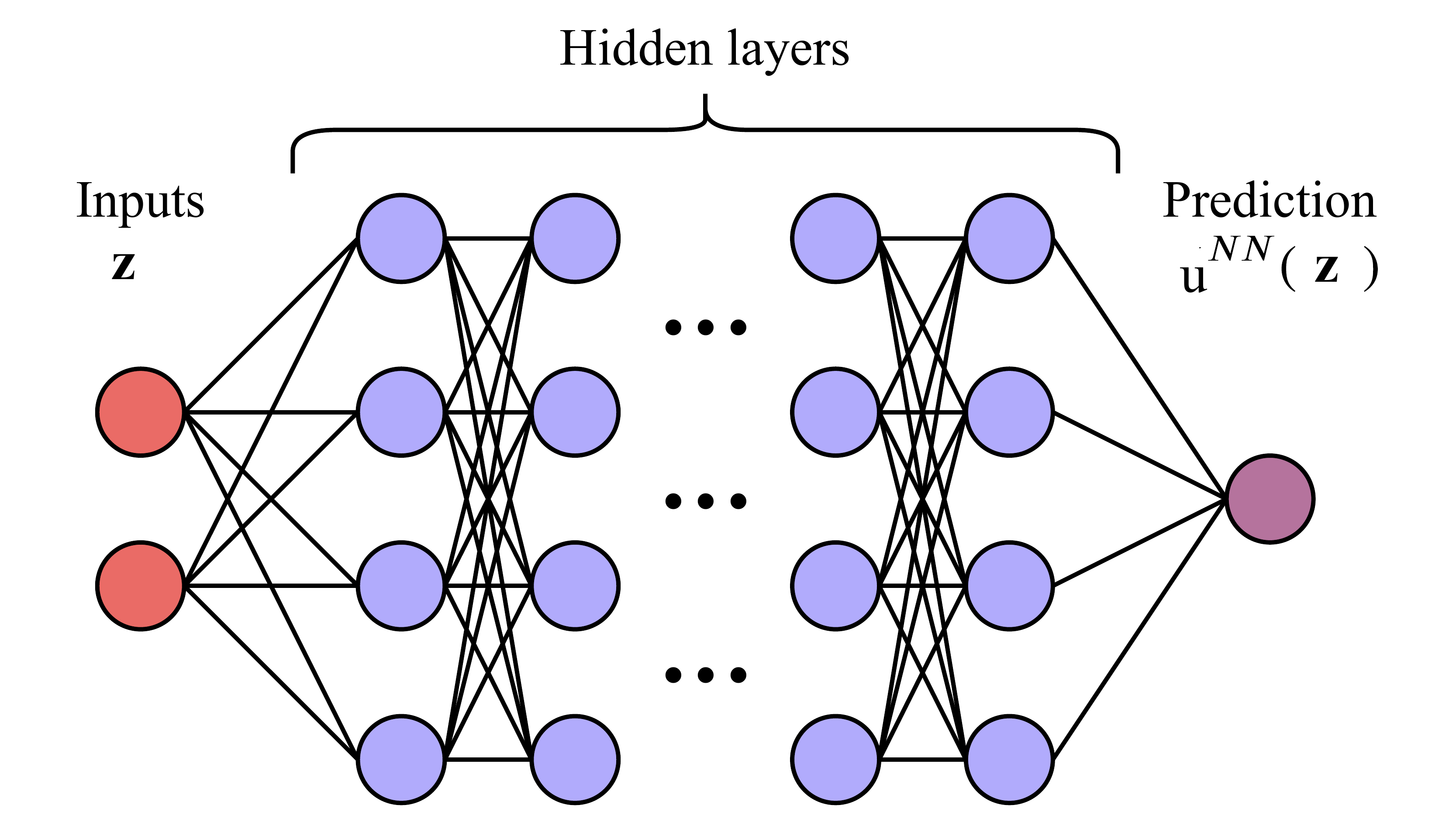}
\caption{ A feedforward neural network }
\label{fig_NNplot}
\end{figure}

In this study, a feedforward neural network (Figure \ref{fig_NNplot}) is a scalar-valued function, $\bfz\in \Real^p \rightarrow u^{NN}\in\Real$
\EQ
\label{eq_NN}
u^{NN}(\bfz) = g_M\circ g_{M-1}\circ \cdots g_1(\bfz)
\EE
where $g_k(\bar\bfz)=\sigma(\bfW_k\bar\bfz +\bfb_k)$, $\bar\bfz$ is a vector (may have different dimensions in different layers of the network), $\sigma$ is a vector-valued activation function such as the hyperbolic tangent, the logistic or ReLU function. The input dimension, $p$, depends on the dimension of $\bfy$ and the number of measurements used in estimation. For example, if $\bfy\in\Real^m$ and if the estimation is based on the information $\{ \bfy(k); 0\leq k\leq K\}$, then $p=m(K+1)$. The value of $u^{NN}$ is an approximation of an individual state variable of the system. In the training process, we minimize the following loss function 
\EQ
l(\bfW,\bfb)=\Fr{1}{\abs{{\cal S}_{training}}}\ds\sum_{(\bfz, u(\bfz))\in {\cal S}_{training}} (u^{NN}(\bfz)-u(\bfz))^2,
\EE
where 
\EQ
\label{eq_S}
{\cal S}_{training}=\left\{ (\bfz, u(\bfz)); \begin{array}{ll} \bfz \mbox{ is in a set of random }\\  \mbox{points in state space}\end{array}\right\}
\EE 
The training is based on the BFGS algorithm to find the parameters $\bfW$ and $\bfb$ that minimize $l(\bfW,\bfb)$. For validation, the accuracy of $u^{NN}(\bfz)$ is evaluated using another data set, ${\cal S}_{validation}$.  

\begin{figure}[!ht]
\centering
\includegraphics[width = 3.0in]{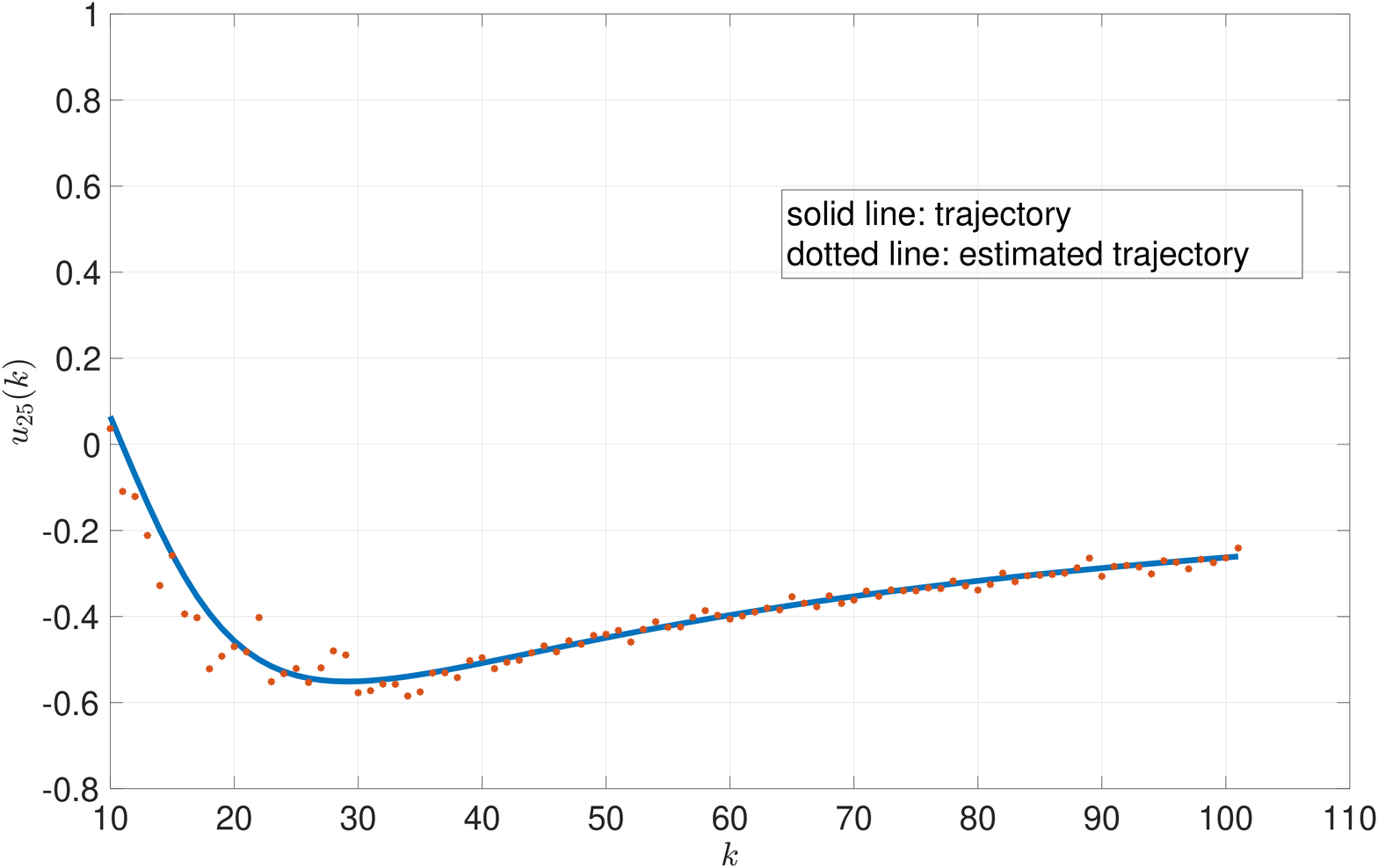}
\caption{ A trajectory and its neural network estimation - Case 1}
\label{fig_NN_a}
\end{figure}

\begin{figure}[!ht]
\centering
\includegraphics[width = 3.0in]{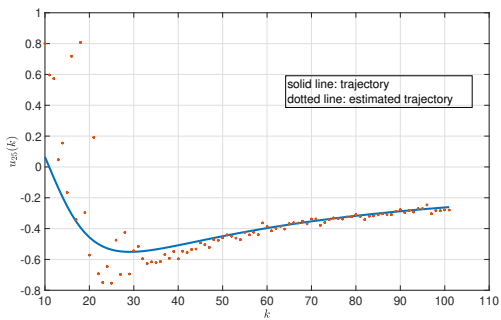}
\caption{ A trajectory and its neural network estimation - Case 2}
\label{fig_NN_b}
\end{figure}

Let us consider the discretized Burgers' equation in Section \ref{sec_UKF} as an illustrative example. We solve the equation using random initial conditions to generate data. Then a neural network is trained so that it can estimate $u_{25}(K)$ based on the sequence of outputs $\{ \bfy(k); 0\leq k\leq K\}$.  The data sets, ${\cal S}_{training}$ and ${\cal S}_{validation}$, are generated using trajectories from random initial states in the following form
\begin{equation}
\begin{array}{r}
\bfu_i(0)= \ds\sum_{j=0}^{N_F}\left(\alpha_j\cos\left(\ds\frac{2\pi j}{L}x_i\right)+\beta_j\sin\left(\ds\frac{2\pi j}{L}x_i\right)\right)\\
i=1,2,\cdots, N_x-1
\end{array}
\end{equation}
subjecting to $\sum \alpha_j =0$ to satisfy the boundary condition (we simply set $\alpha_{N_F}=-\sum_{j<N_F} \alpha_j$ ). The parameters $\alpha_j\sim N(0,\sigma)$ and $\beta_j\sim N(0,\sigma)$ are i.i.d. random variables. In this example, $N_F=3$, $\sigma=0.3$ and $K+1=10$. For each data point in ${\cal S}_{training}$ and ${\cal S}_{validation}$, $\bfz$ is the vector in $\Real^{m(K+1)}$ obtained by reshaping $\{ \bfy(k); 0\leq k\leq K\}$. For each random initial condition and its trajectory, three data points are generated based on $\{ \bfy(s+k); 0\leq k\leq K\}$, where $s=0$ and random integers $s_1$ and $s_2$. For each data point, $u(\bfz)$ is assigned the corresponding value of $u_{25}(s+K)$. The size of the data sets is 
\EQ
\abs{{\cal S}_{training}}=\abs{{\cal S}_{validation}}=3\times 10^4.
\EE
In the validation, the error of $u_{25}^{NN}$ is measured by the root-mean-square error (RMSE)
\EQ
\sqrt{\ds\sum_{(\bfz, u(\bfz))\in {\cal S}_{validation}} \Fr{(u^{NN}(\bfz)-u(\bfz))^2}{\abs{{\cal S}_{validation}}}}
\EE
Two sets of sensor locations are used, 
\EQ
\mbox{Case 1}: \;\bfy(k)=\MT u_{20}(k)& u_{21}(k) & u_{29}(k)& u_{30}(k)\EM^\intercal\\
\mbox{Case 2}: \;\bfy(k)=\MT u_{18}(k)& u_{19}(k) & u_{30}(k)& u_{31}(k)\EM^\intercal.
\EE
Using the first set of sensor locations (Case 1), the averaged observability index of $u_{25}$ over the validation data set is $\rho/\epsilon=4.05$. For Case 2, the number is $43.18$, which indicates that the observability is weaker than Case 1. The trained neural networks have eight layers and $32$ neurons in each layer. The activation function is the hyperbolic tangent. An i.i.d. Gaussian noise is added to $\bfy$. The standard deviation is $0.028$, which is about $5\%$ of the averaged range of $\{ u_{25}(k); 0\leq k\leq K\}$ in the data. The RMSE of $u^{NN}$ is shown in Table \ref{table1}. When there is no sensor noise in $\bfy$, the accuracy of both cases are similar. However, in the presence of sensor noise, the RMSE in Case 1, that has a higher observability, is significantly smaller than Case 2.  A sample trajectory is shown in Figures \ref{fig_NN_a} and \ref{fig_NN_b} for Case 1 and Case 2, respectively. The neural network approximation using sensor locations in Case 2 (dotted line in Figure \ref{fig_NN_b}) has larger error than Case 1 in Figure \ref{fig_NN_a} when $k\leq 30$, the transition phase when the trajectory approaches an equilibrium.  The trained neural network, or the deep filter, provides the approximated trajectory of $u_{25}(k)$ without the need of computing other state variables in $\Real^{49}$. 

\begin{table}[!ht]
\centering
\caption{RMSE of $u_{25}^{NN}$}
\label{table1}
 \begin{tabular}{|c|c|c|c|}
\hline
&&&\\
&$\rho/\epsilon$&&RMSE \\
\cline{1-4}
Sensor location&&noise free&  0.0056\\
\cline{3-4}
Case 1&4.05&with noise&0.0225\\
\hline
Sensor location&&noise free&  0.0066\\
\cline{3-4}
Case 2&43.18&with noise&0.0801\\
\hline
\end{tabular}
\end{table}

%%%%%%%%%%%%%%%%%%%%%%%%%%%%%%%%%%%%%%

\section{Conclusions}
In control theory, the definition of observability based on the observability matrix rank condition is a yes or no answer rather than a quantitative measure. In addition, this definition is about the observability of the overall system state, not applicable to individual state variables. For high dimensional problems, the classical definition of observability does not serve the purpose when estimating the entire system trajectory is either impossible or unnecessary. In this paper,  we introduce a quantitative measure of observability for targeted state variables. For their estimation, we introduce and exemplify a deep filter, which is a neural network specifically designed for the estimation of targeted state variables without computing the trajectory of the entire system. From the examples, both estimations from UKF and deep filter agree with the measure of observability, i.e., the sensor locations that have higher observability (or lower value of the unobservability index) result in more accurate estimation. For future research, more questions are raised than what answered in this paper such as more testing examples that have higher dimensions, finding efficient numerical methods for the quadratic maximization problem that defines the unobservability index, improving the training process for the deep filter, and applying the idea to numerical weather prediction including nowcasting of targeted local areas. \\

\addtolength{\textheight}{-12cm}   % This command serves to balance the column lengths

%%%%%%%%%%%%%%%%%%%%%%%%%%%%%%%%%%%%%%%%%%%%%%%%%%%%%%%%%%%%%%%%%%%%%%%%%%%%%%%%

%%%%%%%%%%%%%%%%%%%%%%%%%%%%%%%%%%%%%%%%%%%%%%%%%%%%%%%%%%%%%%%%%%%%%%%%%%%%%%%%

%%%%%%%%%%%%%%%%%%%%%%%%%%%%%%%%%%%%%%%%%%%%%%%%%%%%%%%%%%%%%%%%%%%%%%%%%%%%%%%%
%\section*{APPENDIX}


%\section*{ACKNOWLEDGMENT AND DISCLAIMER}
\noindent\textbf{Acknowledgment and disclaimer}. This material is based upon activities supported by the National Science Foundation under lnteragency Agreement \#2202668 and Naval Research Laboratory, Monterey, California. Any opinions, findings, and conclusions or recommendations expressed are those of the authors and do not necessarily reflect the views of the National Science Foundation and Naval Research Laboratory.

%%%%%%%%%%%%%%%%%%%%%%%%%%%%%%%%%%%%%%%%%%%%%%%%%%%%%%%%%%%%%%%%%%%%%%%%%%%%%%%%

\end{document}